\documentclass[twocolumn]{aastex63}

\usepackage{units}

\usepackage{amsmath}

\newcommand{\code}[1]{\texttt{#1}}
\newcommand{\mesa}{\code{MESA}}
\newcommand{\MESA}{\mesa}

\newcommand{\Msun}{\ensuremath{\mathrm{M}_\odot}}
\newcommand{\Lsun}{\ensuremath{\mathrm{L}_\odot}}
\newcommand{\Rsun}{\ensuremath{\mathrm{R}_\odot}}
\newcommand{\Msunyr}{\ensuremath{\Msun\,\mathrm{yr^{-1}}}}

\newcommand{\gcc}{\ensuremath{\mathrm{g\,cm^{-3}}}} % density units

\newcommand{\Mch}{\ensuremath{\mathrm{M}_{\rm Ch}}}

\newcommand{\logTeff}{\ensuremath{\log(T_{\rm eff}/\rm K)}}
\newcommand{\logT}{\ensuremath{\log(T/\rm K)}}

\newcommand{\logRho}{\ensuremath{\log(\rho}/\gcc)}

\newcommand{\XC}{\ensuremath{X_{\mathrm{C}}}}
\newcommand{\XO}{\ensuremath{X_{\mathrm{O}}}}

\newcommand{\nuclei}[2]{\ensuremath{\mathrm{^{#1}#2}}}

\newcommand{\neon}[1][20]{\nuclei{#1}{Ne}}

\begin{document}

\author[0000-0002-4870-8855]{Josiah Schwab}
\affiliation{Department of Astronomy and Astrophysics, University of California, Santa Cruz, CA 95064, USA}
\correspondingauthor{Josiah Schwab}
\email{jwschwab@ucsc.edu}

\received{27 July 2020}
\revised{09 October 2020}
\accepted{05 November 2020}

\title{Evolutionary Models for the Remnant of the Merger of Two Carbon-Oxygen Core White Dwarfs}

\begin{abstract}
  We construct evolutionary models of the remnant of the merger of two
  carbon-oxygen (CO) core white dwarfs (WDs).  With total masses in
  the range $\unit[1-2]{\Msun}$, these remnants may either leave
  behind a single massive WD or undergo a merger-induced collapse to a
  neutron star (NS).  On the way to their final fate, these objects
  generally experience a $\sim \unit[10]{kyr}$ luminous giant phase,
  which may be extended if sufficient helium remains to set up a
  stable shell-burning configuration.  The uncertain, but likely
  significant, mass loss rate during this phase influences the final
  remnant mass and fate (WD or NS).  We find that the initial CO core
  composition of the WD is converted to oxygen-neon (ONe) in remnants
  with final masses $\gtrsim \unit[1.05]{\Msun}$.  This implies that
  the CO core / ONe core transition in single WDs formed via mergers
  occurs at a similar mass as in WDs descended from single stars, and
  thus that WD-WD mergers do not naturally provide a route to
  producing ultra-massive CO-core WDs.  As the remnant contracts
  towards a compact configuration, it experiences a ``bottleneck''
  that sets the characteristic total angular momentum that can be
  retained.  This limit predicts single WDs formed from WD-WD mergers
  have rotational periods of $\approx \unit[10-20]{min}$ on the WD
  cooling track.  Similarly, it predicts remnants that collapse
  can form NSs with rotational periods $\sim \unit[10]{ms}$.
\end{abstract}


\section{Overview}

The merger of two white dwarfs (WDs) can have a range of outcomes
depending on the masses and compositions of the WDs
\citep[e.g.,][]{Webbink1984, Iben1985a}.  The merger of a He WD with another He
WD or low mass CO WD re-initiates stable He burning and the resulting
merged object spends nuclear timescales
$(\sim \unit[10^5 - 10^8]{yr})$ as a hot subdwarf or R CrB star,
before eventually going down the cooling track as a single WD
\citep[e.g.,][]{Schwab2018, Schwab2019}.
In the case of mergers involving more massive WDs, the focus has
primarily been on systems where the merger is likely to promptly lead
to an explosive transient like a Type Ia supernova
\citep[e.g.,][]{Shen2018b, Perets2019}.  Our current theoretical
understanding does not definitively map the WD masses at merger to the
set of possible final outcomes.  However, it seems likely that there
is at least some subset of WD-WD mergers with total mass
$\gtrsim \unit[1]{\Msun}$ that do not immediately destroy the system
(i.e., the primary WD does not detonate).

In the non-destructive case, this leaves behind a
$\approx~\unit[1-2]{\Msun}$ CO-dominated merger remnant.  Broadly, the
outcome is expected to be dependent on the remnant mass, with systems
below the Chandrasekhar mass ($\Mch$) producing massive single WDs and
systems above $\Mch$ undergoing a merger-induced collapse (MIC) to form a
neutron star \citep[NS; ][]{Nomoto1985, Saio1985b}.
\citet{Schwab2016b} demonstrated that this latter process may proceed
through the formation of a low-mass iron core, evolving similarly to
the exposed, low-mass metal cores found in the progenitors of
ultra-stripped supernovae \citep[e.g.,][]{Tauris2015b}.

Ongoing observational developments motivate understanding the
signatures of WD-WD mergers in this mass range.
Recent work demonstrates that there is class of massive
($\approx \unit[0.8-1.3]{\Msun}$) DQ WDs with distinguishing chemical
and kinematic properties \citep{Dunlap2015, Coutu2019, Koester2019,
  Cheng2019}.  These objects have been suggested to be WD-WD
merger remnants, but seem too He-poor and too massive to be the
descendants of the R CrB stars.
Further kinematic analysis of \textit{Gaia} data suggests
$\approx 20\%$ of all massive WDs may be merger products
\citep{Cheng2020a}.
Other peculiar individual objects have emerged.
\citet{Hollands2020} report a massive WD
($\approx \unit[1.15]{\Msun}$) with an unusual carbon-hydrogen
atmosphere and fast kinematics, potential signatures of a merger.
\citet{Gvaramadze2019b} report the detection of a hot, luminous object
in a H-and-He-free nebula that roughly resembles the predictions of
\citet{Schwab2016b} for the properties of a double CO WD merger
remnant $\sim 10$ kyr post-merger.

A theoretical understanding of the evolution of WD-WD merger
remnants begins with knowledge of the conditions that develop in the
immediate aftermath of the dynamically-unstable mass transfer in
that can occur in these double-degenerate binaries.  Modern insight
came via smoothed-particle hydrodynamics (SPH) simulations by
\citet{Benz1990} and has been refined over the subsequent decades.
Beginning from this post-merger configuration, one must then follow
the remnant into its longer-duration phases.  In an early milestone,
\citet{Segretain1997} created hydrostatic models that resembled
merged configurations and used them to suggest the likely importance
of the loss of mass and angular momentum from the remnant.  In a
pioneering work, \cite{Yoon2007} modeled WD-WD merger remnants by
performing stellar evolution calculations beginning from initial
conditions based on the results of WD-WD merger simulations.  Much
subsequent work, including this paper, follows in that vein.

Here, we construct simple evolutionary models of WD-WD merger
remnants with (initially) CO cores and total masses
$\gtrsim \unit[1]{\Msun}$. Our goal is to investigate their final
fates and describe the effects of key physical processes on the
post-merger evolution.  Section~\ref{sec:models} describes how we
construct our initial conditions and Section~\ref{sec:post-merger}
describes the baseline post-merger evolution.
Sections~\ref{sec:mass-loss}, \ref{sec:nuclear-burning}, and
\ref{sec:rotation} discuss the additional effects of mass loss,
nuclear burning, and rotation, respectively.
In Section~\ref{sec:conclusions} we summarize and conclude.

\section{Models}
\label{sec:models}

We construct stellar evolution models of the merger remnants using
\MESA\ r12778 \citep{Paxton2011, Paxton2013, Paxton2015, Paxton2018,
  Paxton2019}.%
\footnote{Our input files are publicly available at \url{https://doi.org/10.5281/zenodo.4075491}.}
The initial models are guided by the post-dynamical-phase structures
from \citet{Dan2014}, but we do not attempt to directly map these
results into \MESA.  Instead, we create a set of parameterized initial
\MESA\ models that reflect the key features of the remnant structure
and schematically include the effects of the viscous phase
\citep{Shen2012, Schwab2012}.  We first describe the procedure that we
use to construct non-rotating, pure carbon-oxygen models and then
later describe how we extend these models to explore the effects of
rotation and the presence of helium.

\subsection{Initial Conditions}
\label{sec:ics}

\citet{Dan2014} performed a large set of WD merger simulations that
span the space of mass/composition for each of the primary and
secondary WD.  Thermodynamic and rotational profiles of their models
were previously available
online.\footnote{The link given in their paper became inactive while this work was being performed.  We will include their data that we previously accessed and used along with our input files.}
Their HeCO WDs have masses 0.50, 0.55, and 0.60 \Msun\ and have pure
CO cores ($\XC = \XO = 0.5$) overlaid by a $\unit[0.1]{\Msun}$ pure-He
mantle.  Their CO WDs have masses $0.65-1.05$ \Msun\ and are pure
carbon-oxygen ($\XC = 0.4, \XO = 0.6$) throughout.  In this work, our
default assumption for the composition of the models is pure CO
($\XC = 0.4, \XO = 0.6$), but Section~\ref{sec:nuclear-burning}
discusses the effects of the presence of He.

The \citet{Dan2014} calculations follow the
merger through its dynamical phase---covering three orbital periods % ($\sim$ minutes)
after the donor is disrupted---and result in the formation of a
roughly axisymmetric remnant.  \citet{Shen2012} emphasize the
importance of a subsequent viscous phase of evolution, lasting for
$\sim \unit[10^4-10^8]{s}$, during which angular momentum (AM) transport
due to magnetohydrodynamic (MHD) processes brings the remnant toward
solid-body rotation long before significant cooling can occur.  The
effects of this phase are important to include, but are challenging to
follow with numerical simulations \citep[e.g.,][]{Schwab2012, Ji2013}.

The information made publicly available by \citet{Dan2014} are
1D profiles (slices and equipotential averages) of their
simulations.  These are not sufficient to directly initialize
multi-D hydrodynamic simulations of the viscous phase like those
performed in
\citet{Schwab2012} and \MESA\ does not have appropriate capabilities
for approximately following the viscous phase in 1D.  As such, we
elect instead to schematically include the effects of the viscous
phase (described in more detail below).  Future work directly
modeling the viscous phase can eliminate some of the uncertainties
introduced by our schematic approach.

\citet{Dan2014} divide the remnant into 4 components: the core, hot
envelope, disk, and tidal tail.  We use their provided fitting
formulae (which are functions of total mass and mass ratio) to set the
masses of each of these components.  We identify each of our
models primarily by its mass ratio and total mass (i.e., ``the
$q=2/3$, $M_{\rm tot} = 1.5\,\Msun$ model'').

The tidal tail accounts for at most a few percent of the total mass
and we therefore ignore it.  We neglect any mass unbound from the
system (so that the initial remnant mass remains the same as the total
mass of the merging WDs) and do not attempt to include the effect of
tidal tail material that may remain bound and fall back.  However, we
do note that this cool tidal tail material is a potential reservoir of
unburned H or He.  This material could then be incorporated into the
outer layers of the remnant as the remnant expands and/or the tidal
material returns from apocenter.  This might provide an avenue for
creating thin H/He layers on merger remnant WDs.

The core is the inner portion of the primary WD.  Our initial model
has an isothermal core of mass $M_{\rm core}$ with temperature
$T_{\rm core}$.  The precise temperature of this material is generally
unimportant, so long as it is degenerate.  We typically assume
$T_{\rm core} \approx \unit[10^8]{K}$.

The hot envelope is material that was shock heated in the merger.
This is exterior to the core and includes the initial outer layers of
the primary.  Because the viscous evolution will also transform the
disk material into a hot envelope, we instead refer to this region of
the model as the ``peak'', since it contains the temperature peak.
This region has a mass $M_{\rm peak}$ with a maximum temperature
$T_{\rm peak}$.  We assign this region an entropy profile that
linearly increases with mass coordinate and connects the low-entropy
core and the high-entropy disk.

\begin{figure}
  \centering
  \includegraphics[width=\columnwidth]{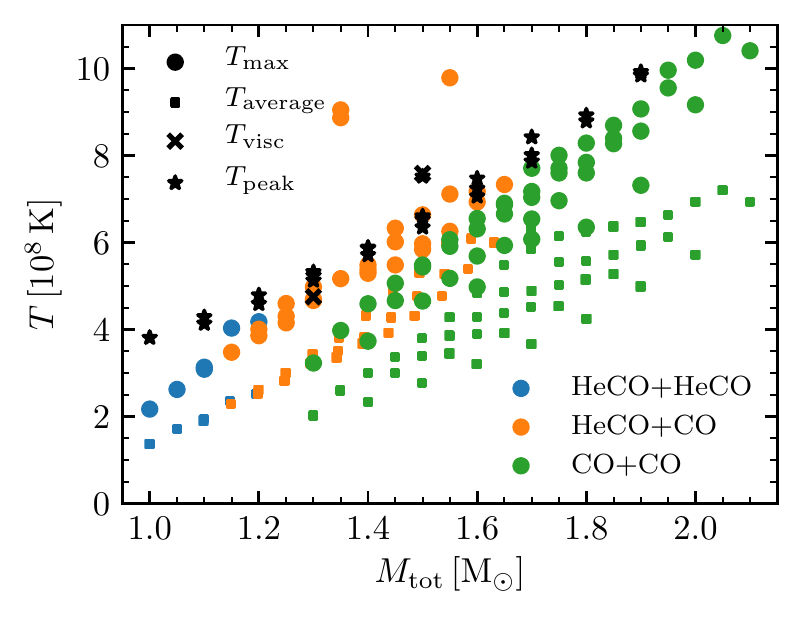}
  \caption{Maximum temperature versus total mass of the merging WDs
    from \citet{Dan2014}.  Larger circles show the maximum SPH
    particle temperature while smaller squares show the maximum value
    from equipotential-averaged profiles.  Colors indicate the
    compositions of the merging WDs.  The Xs indicate the maximum
    temperature of two models after simulations of the viscous evolution (see text).
    The stars mark the peak temperatures of the initial models used in
    this work.}
  \label{fig:Tfit}
\end{figure}

Figure~\ref{fig:Tfit} shows the $T_{\rm max}$ values reported in
\citet{Dan2014} as a function of the total mass of the merging WDs.
The value of $T_{\rm max}$ is the maximum over the simulation and so
typically reflects a localized hot region. The squares in
Figure~\ref{fig:Tfit} show the maximum temperature extracted from the
equipotential-averaged, 1D profiles of the \citet{Dan2014} models.
These temperature values are systematically lower than $T_{\rm max}$
and would more accurately reflect the temperatures implied by a direct
mapping of the post-dynamical phase structure into (1D) \MESA.
However, in the viscous calculations of \citet{Schwab2012}, the peak
temperature increases relative to its value at the end of the
dynamical phase.  This is primarily due to adiabatic compression from
material above as the rotational support of the disk material is
removed (though in some cases additional entropy can be injected due
to nuclear burning).
Therefore, by using the higher $T_{\rm max}$, we are assuming that
these two differences approximately offset.  As a check of this crude
assumption, the Xs in Figure~\ref{fig:Tfit} mark the final maximum
temperature from the end of two viscous-phase
calculations.\footnote{The model with
  $M_{\rm tot} = \unit[1.5]{\Msun}$ (0.6+0.9) is the viscous evolution
  model ZP5c from \citet{Schwab2012}. See Figures 1-5 in that
    work for a more detailed illustration of the effects of the viscous
    phase.  The model with
  $M_{\rm tot} = \unit[1.3]{\Msun}$ (0.65+0.65) is an unpublished
  simulation using the same methods.} On the basis of this limited
check, there is no indication that this choice is dramatically
incorrect.

The disk material is primarily the tidally disrupted secondary and is
initially rotationally supported.  Viscous dissipation subsequently
heats this material as its angular momentum is transported outward.
\citet{Schwab2012} found that this region became convectively unstable
and thus evolved towards an entropy profile that is roughly spatially
constant.  Therefore, we assign this material such a state.
Because we do not directly simulate the viscous phase, we do not
know the precise entropy to target, so instead we select a
total-mass-dependent entropy that gives us the desired trend in
$T_{\rm peak}$.
The state of the outer layers is the most artificial aspect of our
treatment.  This limits the predictive power of our models in their
earliest phases.  However, after the first few thermal times of this
envelope elapse, we expect its state to be reset by the luminosity from the
hot material below.

In practice, our prescription is as follows. We begin with a high entropy carbon-oxygen \MESA\
model of the desired mass and relax its temperature/entropy following
a procedure similar to that described in Appendix~A of
\citet{Schwab2016b}.  Our target profile is:
% \begin{equation}
% \begin{cases} 
%       T(M_r) = T_{\rm core} & 0 \le M_{r} \le M_{\rm core} \\
%       s(M_r) = s_{\rm core} + [s_{\rm env} - s(M_{\rm core})] \frac{M_{r} - M_{\rm core}}{M_{\rm peak}} &
%       M_{\rm core} \le M_{r} \le M_{\rm core} + M_{\rm env} \\
%       s(M_r) = s_{\rm env}  & M_{\rm core} + M_{\rm env} < M_{r} \le M_{\rm tot}\\
%    \end{cases}
%  \end{equation}
for $0 \le M_{r} \le M_{\rm core}$,
\[T(M_r) = T_{\rm core}~;\]
for $ M_{\rm core} \le M_{r} \le M_{\rm core} + M_{\rm env}$,
\[s(M_r) = s_{\rm core} + [s_{\rm env} - s(M_{\rm core})] \frac{M_{r} - M_{\rm core}}{M_{\rm peak}}~;\]
and for $M_{\rm core} + M_{\rm env} < M_{r} \le M_{\rm tot}$,
\[s(M_r) = s_{\rm env}~.\]
We adopt $T_{\rm core} = \unit[10^8]{K}$ and the ad hoc
relationship
$\log(s_{\rm env}/\rm erg\,g^{-1}\,K^{-1}) = 8.7 + 0.3 \left(M_{\rm
    tot}/\Msun - 1.5\right)$ in order to achieve the desired
$M_{\rm tot}$ vs. $T_{\rm peak}$ relationship shown in
Figure~\ref{fig:Tfit}.  Given this somewhat arbitrary form, in
Section~\ref{sec:post-merger} we show models at a fixed mass with
varying values of $s_{\rm env}$ to illustrate that the evolutionary
trajectory is not sensitive to this choice.

\begin{figure}
  \centering
  \includegraphics[width=\columnwidth]{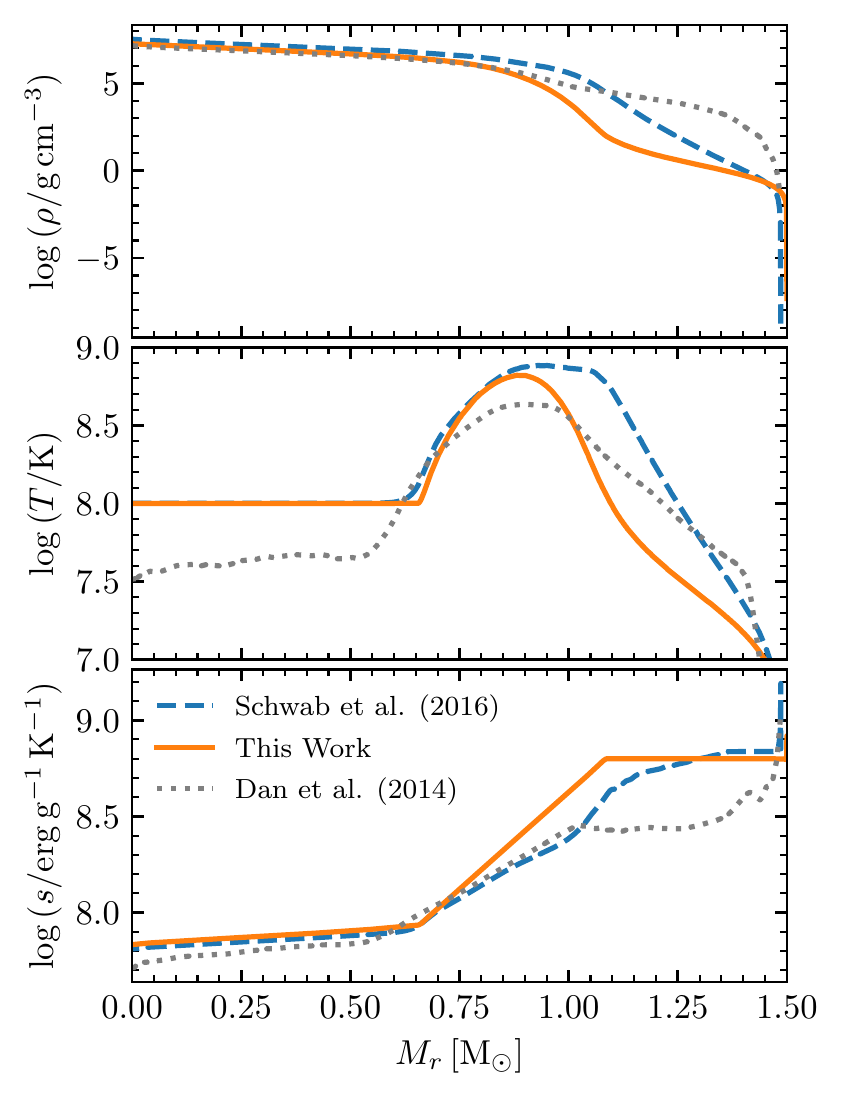}
  \caption{Thermodynamic profiles of the parameterized initial model
    used in this work (solid lines).  For comparison, we show (dashed lines) the equivalent initial
    model from \citet{Schwab2016b}, which was based on a higher
    resolution SPH simulation from \citet{Dan2011} and a simulation of
    the viscous phase from
    \citet{Schwab2012}\label{fig:ics-profiles}.
    We also show (dotted lines) the 1D equipotential-averaged,
    post-dynamical-phase profiles from \citet{Dan2014}.
  }
\end{figure}

In Figure~\ref{fig:ics-profiles}, we demonstrate that this approach
provides a reasonable starting condition.  We compare with the
post-viscous-phase model of a $\unit[0.6 + 0.9]{\Msun}$ CO+CO WD
merger from \citet{Schwab2012} as the detailed entropy profile of this
model provided the initial condition for the fiducial model in \citet{Schwab2016b}.
% see also Figure~A1 in that paper..
The schematic model reproduces the key features, namely a hot envelope
overlaying a cold core.  The dotted lines show the 1D
equipotential-averaged, post-dynamical-phase profiles from
\citet{Dan2014}.  The differences from the post-viscous-phase model
above the core (at $M_r \gtrsim \unit[0.65]{\Msun}$) illustrate the
increase in the peak temperature and the envelope entropy that occur
during the viscous phase.  As it was designed to do, the schematic approach
provides an initial condition that is a closer match to the peak
temperature and envelope entropy of the more detailed
post-viscous-phase model than a model beginning directly from a
\citet{Dan2014} profile.

\begin{figure}
  \centering
  \includegraphics[width=0.49\textwidth]{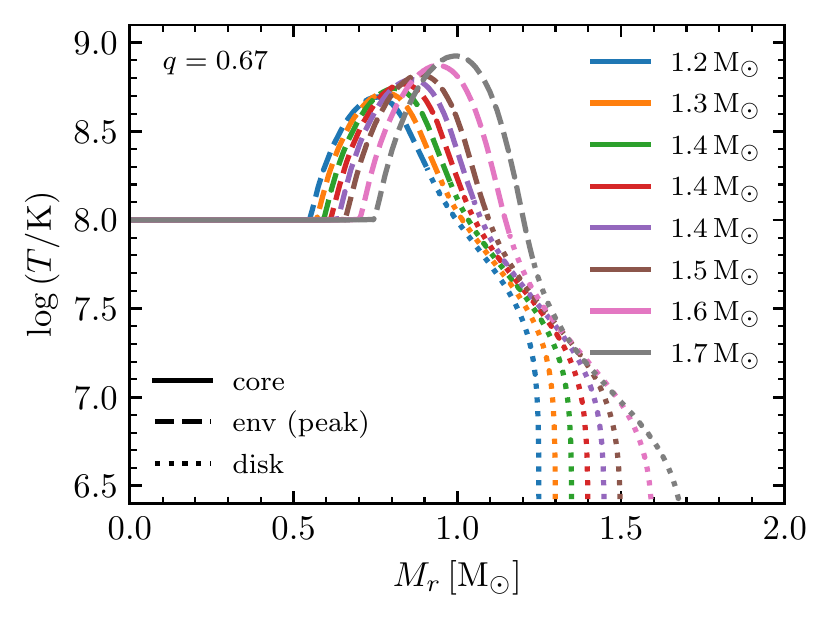}
  \includegraphics[width=0.49\textwidth]{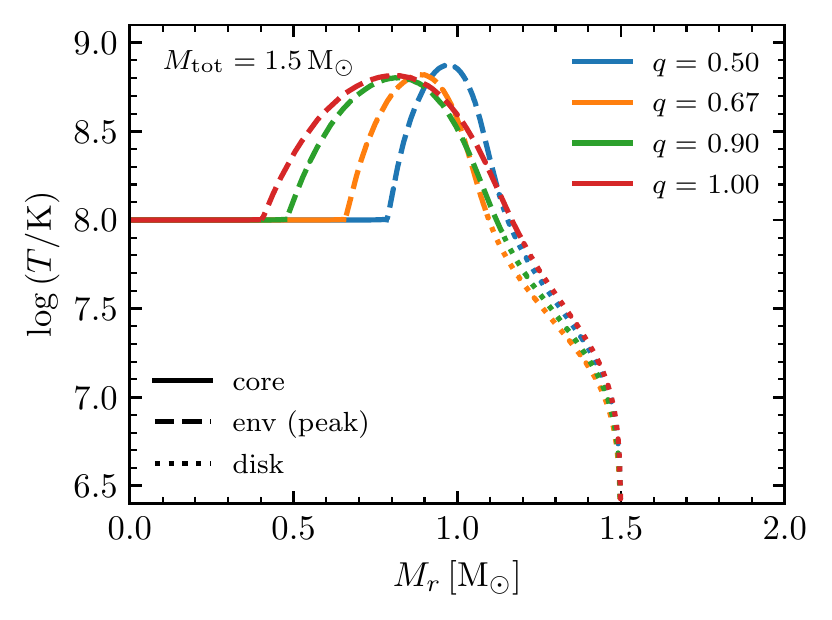}
  \caption{Temperature profiles of families of parameterized initial
    models.  The top (bottom) panel shows a sequence of models at
    fixed (varying) mass ratio and varying (fixed) total mass.  The
    line style indicates the different components following the
    \citet{Dan2014} subdivision. \label{fig:ics-variation}}
\end{figure}

While lacking some of the consistency that would come from a more elaborate viscous-phase simulation of the merger,
the simple parameterized nature of these initial conditions allows us
to easily generate families of models that vary the total mass and
mass ratio. Figure~\ref{fig:ics-variation} illustrates how the
temperature profiles vary in two such families.

\subsection{Input Physics}

We use the OPAL radiative opacities for C- and O-rich mixtures
\citep{Iglesias1993, Iglesias1996}, referred to as ``Type 2'' tables
in \MESA, with a base metallicity $Z = 0.02$.  The stellar models
evolve to temperatures below the lower boundary of the OPAL
tabulations ($\logT < 3.75$), so we supplement this with the
low-temperature table generated in \citet{Schwab2016b}---see their
Appendix B---that smoothly extends the opacities to lower
temperatures.

The \MESA\ equation of state (EOS) compilation does not include any
component EOS that covers CO mixtures and includes ionization of these
metals.  Therefore, in the regions [$\logT \lesssim 7.7$ and
$\logRho \lesssim 3.5$] that would normally be covered by the OPAL
and/or PTEH EOSes when $Z < 1$, we instead use the HELM EOS
\citep{Timmes2000b}.  HELM includes an ideal gas of ions, a
Fermi-Dirac electron gas, and radiation.  It parameterizes composition
by the mean ion weight and charge.  HELM is a physically suitable
choice for when material is fully ionized and is also the EOS used in
the merger simulations of \citet{Dan2011, Dan2014}.  The default
behavior in \MESA\ is to crudely mock up the effects of ionization by
blending between versions of HELM including the contribution of
electrons over the temperature range $\logT = 4.5 - 5.0$.  In \MESA\
r6596, as used in \citet{Schwab2016b}, this blend was hard coded in.
In \MESA\ r12778, this is user-configurable, and in this work, we
deactivate this blend and so use an EOS assuming full ionization
throughout.
We use the 21 isotope, $\alpha$-chain nuclear
network \texttt{approx21.net}.

As the remnant expands towards a giant structure, it develops
radiation-pressure-dominated envelope that is locally super-Eddington.
Convection, as modeled by mixing length theory (MLT), becomes
inefficient and a steep entropy gradient develops at the base of the
convective region.  Tracking this narrow region is numerically
demanding and often severely limits the timestep.  To circumvent this,
we apply the ad hoc ``MLT++'' prescription discussed in
\citet{Paxton2013}.  This procedure essentially assumes a third
(unspecified) pathway for energy transport, allowing the temperature
gradient to be closer to adiabatic and the entropy gradient less steep.

Taken together, the limited opacities, lack of appropriate EOS, and
and use of MLT++ mean that the outer layers of our stellar models are
poorly modeled.  The luminosity is typically set by the energy
transport in the deeper, better-modeled layers, but the
radii/effective temperatures of our models are best understood as
qualitative predictions.

\subsection{Stopping Conditions}

We evolve the models until they either begin to go down the WD cooling
track (stopping when $L < \unit[100]{\Lsun})$ or experience off-center
Ne ignition.  For those models that experience Ne ignition, we are
unable to follow the models further due to the computational expense
of resolving the thin neon-oxygen-burning flames that develop.  (The
one such calculation reported in \citet{Schwab2016b} took months to
run.)  However, because the Ne/O-burning flames are relatively fast
$(\sim 10\,\rm cm\,s^{-1}$) and subsequent Si-burning flames even
faster $(\sim 10^3\,\rm cm\,s^{-1}$), the timescales for these burning
stages are $\sim \unit[10]{yr}$ and $\sim \unit[0.1]{yr}$ respectively
\citep{Timmes1994, Woosley2015}.  As such, while the core continues to
evolve in response to these burning processes, we expect that the
outer layers are approximately frozen at this time.  Therefore, while
we do not model it, we expect the remnants that experience Ne ignition
and are above \Mch\ to undergo a MIC to form a NS
via the formation of a low mass Fe core as outlined by
\citet{Schwab2016b}.  Those few remnants with masses sufficiently high
to ignite Ne but that remain below \Mch\ may instead leave behind massive
single WDs with Si-group or Fe-group core compositions.

\subsection{Comparison with \citet{Schwab2016b}}

\begin{figure}
  \centering
  \includegraphics[width=\columnwidth]{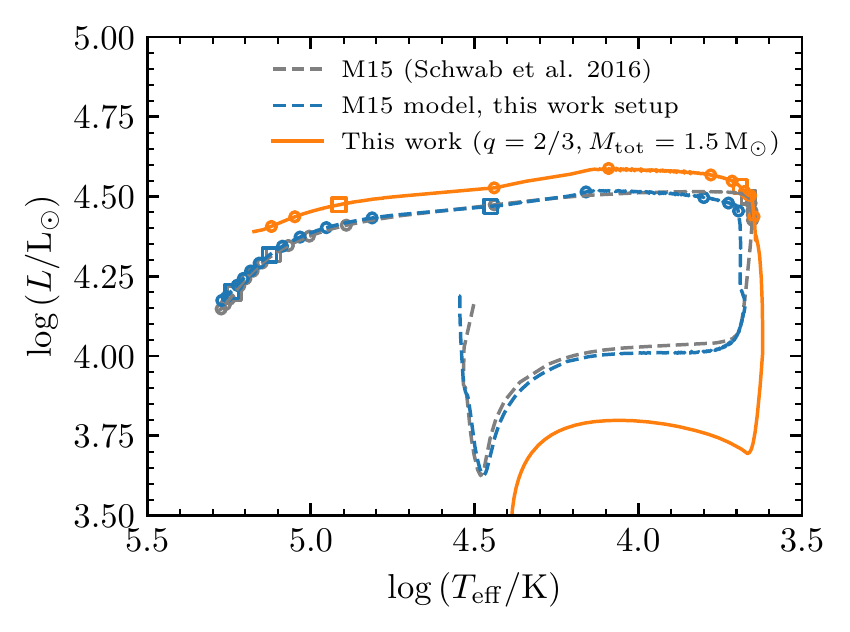}
  \caption{Comparison of the evolution of 0.6\,\Msun\ + 0.9\,\Msun\
    merger models in the HR diagram.  These models stop when carbon
    burning reaches the center of the remnant.  Time moves
    counterclockwise along the tracks with small circles each kyr and
    larger squares every 5 kyr. The two dashed lines illustrate the
    small differences between model M15 from \citet{Schwab2016b} and
    the result of this same initial model evolved using the \MESA\
    version and options adopted in this work.  The solid line shows
    the result of the parameterized initial model adopted in this
    work, illustrating the differences in evolution due to the initial
    model differences shown in
    Figure~\ref{fig:ics-profiles}. \label{fig:S16-HR}}
\end{figure}

To compare the \MESA\ setup used in this work with that in
\citet{Schwab2016b}, we evolved the same fiducial initial model
``M15''.  Figure~\ref{fig:S16-HR} shows the model track in the HR
diagram from shortly after the merger until carbon burning reaches the
center.  The two M15 model tracks agree closely, though the time
evolution is slightly different with the current setup, resulting in a
model that spends less time in the reddest part of the HR diagram but
moves to the blue more slowly.  The analogous parameterized model
shows more substantial differences, though these simply reflect its
different initial structure.

\section{Post Merger Remnants}
\label{sec:post-merger}

To illustrate how the evolution depends on the properties of the
merging binary, we generate families of initial models following the
scheme described in Section~\ref{sec:ics}.  We focus on masses
$M_{\rm tot} \approx \unit[1-2]{\Msun}$ with mass ratios such that
neither of the component WDs would likely be He WDs $(< 0.5\,\Msun)$
or ONe WDs $(> 1.05\,\Msun)$.

\begin{figure*}
  \centering
  \includegraphics[width=0.49\textwidth]{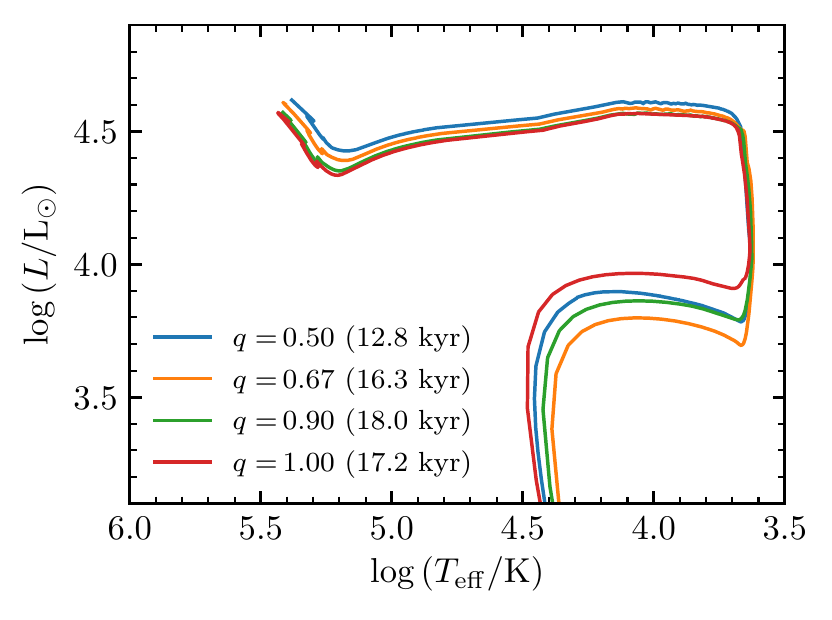}
\includegraphics[width=0.49\textwidth]{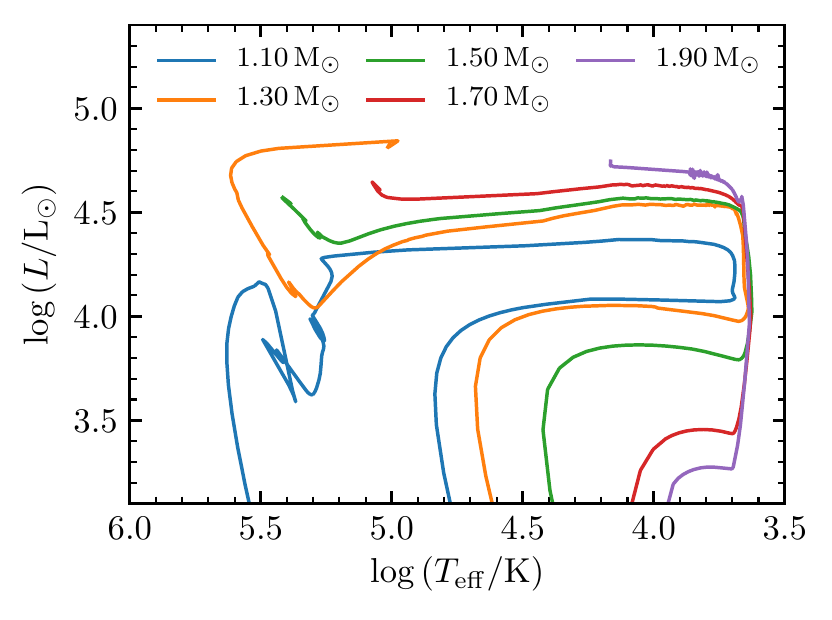}
\caption{Model evolution in the HR diagram.  Models stop when they
  undergo off-center Ne ignition or start to go down the WD cooling
  track.  The left panel shows a sequence at fixed total mass
  $(M_{\rm tot} = 1.5\,\Msun)$ and the legend indicates the duration
  of the plotted track.  The right panel shows a sequence of models at
  fixed mass ratio $(q = 0.9)$.  The green line is the same in the two
  plots.  The total mass of the remnant is the most important
  parameter controlling its evolution. \label{fig:HR-variation}}
\end{figure*}

Figure~\ref{fig:HR-variation} shows the evolution of some of these
models in the HR diagram.  The left panel shows a sequence at fixed mass
$(M_{\rm tot} = 1.5\,\Msun)$ and varying mass ratio.  All these models
ignite carbon burning and terminate at Ne ignition.  The HR evolution
is similar and the duration of the evolution (indicated in legend) is
also within $50\%$.  Given the level to which we trust our input
physics and the details of the initial models, we regard these tracks
as effectively identical.  These models do not predict a significant
dependence on the mass ratio, where the main mass-ratio-dependent
property is the initial location and width of the temperature peak
region.

The right panel shows a sequence of models at fixed mass ratio $(q = 0.9)$ 
and varying total mass.  All models experience carbon ignition.  The
1.1 \Msun\ model becomes a ONe WD.  The 1.3 \Msun\ model experiences
numerical problems after the C flame reaches the center.  This appears
to be associated with a high luminosity during the subsequent KH
contraction phase (around its bluest extent).  If this model were
allowed to launch a wind in response, we expect that it would shed its
outer layers (the source of the numerical issues) and become a massive
ONe WD. The 1.5, 1.7, and 1.9 \Msun\ models all reach Ne ignition.
The models evolve more rapidly with increasing mass (18, 13, and 10
kyr respectively), with this most massive model reaching Ne ignition
while it still has a relatively extended envelope
$(R \approx 30\,\Rsun)$.

The luminosity roughly scales with the mass, being near-Eddington.
The small dynamic range in expected remnant masses (at most a factor
of 2) makes this likely to be challenging as an observational
diagnostic, but our models do generically predict that more massive
remnants are more luminous.

\begin{figure}
  \centering
\includegraphics[width=\columnwidth]{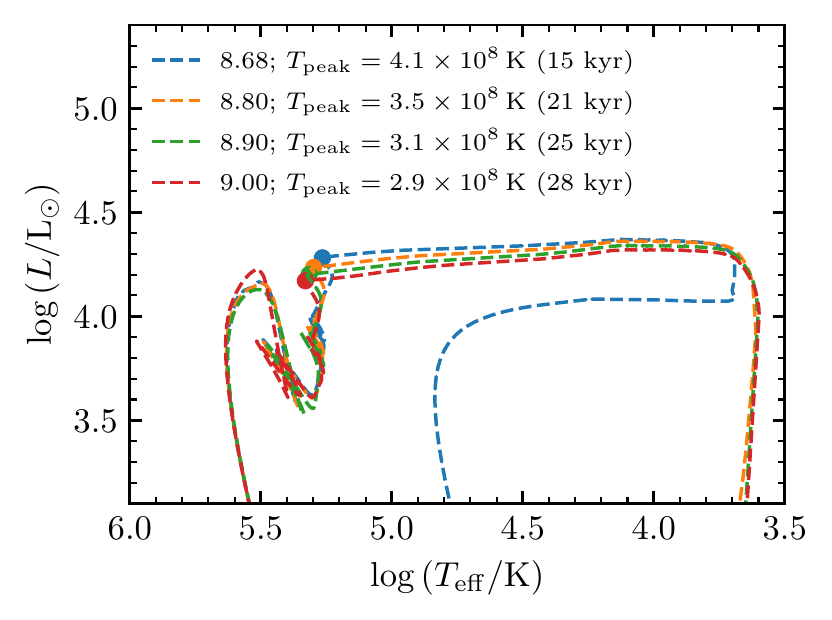}
\caption{Comparison of evolution in the HR diagram for the $q=0.9$,
  $M_{\rm tot} = 1.1\,\Msun$ model with varying values of the initial
  envelope entropy.  The first number in each legend entry indicates
  the value $\log(s_{\rm env}/\rm erg\,g^{-1}\,K^{-1})$.  This is
  followed by the initial peak temperature and the time from the start
  of the calculation until carbon is ignited (which is indicated by
  the solid dot). The overall evolution is not strongly influenced by
  this aspect of the initial condition.  \label{fig:HR-senv}}
\end{figure}

The value of the parameter $s_{\rm env}$ in our models was set in an
ad hoc way.  In Figure~\ref{fig:Tfit}, the peak temperatures in our
remnant models with masses $\lesssim \unit[1.2]{\Msun}$ are somewhat
higher than those suggested by the merger calculations.  Raising the
envelope entropy will lower the peak temperature.  Since the temperatures are
already well below the $\sim \unit[10^9]{K}$ associated with carbon
burning, the initial nuclear energy release will be negligible and so
we would not expect our results to be sensitive to modest variations in
$T_{\rm peak}$.  Figure~\ref{fig:HR-senv} illustrates that varying
$s_{\rm env}$ has only a minor effect on the evolution.  The higher
entropy models are puffier and so begin as giants on the HR diagram.
Given the higher entropy, it also takes longer for the envelopes to
cool and contract.  The legend indicates the length of time between
the start of the calculation and the initiation of carbon burning
(marked by the dot).

\section{Effects of Mass Loss}
\label{sec:mass-loss}

The thermal energy deposited as a result of the merger causes the
remnant to inflate to giant radii ($\gtrsim \unit[100]{\Rsun}$).
Given the CO-dominated composition of the envelope, it seems likely
that these objects will drive dusty winds.%
\footnote{The C-rich R CrB stars \citep[e.g.,][]{Clayton2012}---thought to be He WD + CO WD merger remnants---have observed dust formation/ejection events and are surrounded by dusty shells \citep{Montiel2015b, Montiel2018}.  The possible merger remnant reported by \citet{Gvaramadze2019b} also exhibits a shell of material around the central object.}
The mass loss of such
objects is not observationally well-constrained and we do not have a
reliable way of theoretically estimating these mass loss rates.  While
the mass loss rate will be important for setting the observed
properties of the remnant, the overall evolution---in particular the
extent to which the object experiences C-burning and Ne-burning (and
beyond)---is also influenced by the total amount of mass lost.

We explore the effect of mass loss using an ad hoc wind prescription.
Given that the time spent as a giant is $ \sim 10$ kyr, only mass loss
rates $\gtrsim 10^{-5}\,\Msunyr$ can have a significant evolutionary
effect by changing the total mass.  We assume that the object will
shed a fraction of its mass $f$ on
a timescale $\frac{GM^2}{RL}$, evaluated at the surface.%
\footnote{The timescale $\frac{GM^2}{RL}$ at the surface is $\sim 10$ yr.  This is much shorter than the $\sim 10$ kyr giant phase duration, which is why
significant giant phase mass loss corresponds to $f \sim 10^{-4} \ll 1$.
The $\sim 10$ kyr reflects the timescale for
the object to radiate the energy contained in the thermally-supported material above the degenerate core.  That reservoir is at a much smaller radius than the surface, $\sim \unit[0.03]{\Rsun}$, and so has much higher specific energy than the surface material.}
%
% In the models, there is ~0.5 Msun of material supported by thermal
% pressure at R ~ 1e9-1e10 cm, and so with a specific energy cpT ~ GM/R
% ~ 1e17 erg/g.  Then radiating this at L ~ 3e4 Lsun takes ~ 10 kyr.
%
This implies a mass loss rate
\begin{equation}
\begin{split}
  \dot{M} = 10^{-5}\,\Msun\,{\rm yr}^{-1} & \left(\frac{f}{10^{-4}}\right)
  \left(\frac{R}{100\,\Rsun}\right)\\
  \times & \left(\frac{L}{3\times10^4\,\Lsun}\right)
  \left(\frac{M}{\Msun}\right)^{-1},
  \label{eq:mdot-kh}
\end{split}
\end{equation}
where our fiducial value of $f$ is selected to give
$\dot{M} = 10^{-5}\,\Msunyr$.  Compared to an even simpler form like a
constant, Equation~\eqref{eq:mdot-kh} has the advantage that the mass
loss rate is smaller when the object is compact and larger when the
object is a luminous giant.  When discussing models using this
prescription, we will indicate the value of $f$ using the shorthand
$f_{-4} \equiv f/10^{-4}$.  This prescription is equivalent to the
\citet{Reimers1975} mass loss prescription for red giants with the
scaling factor $\eta_{\rm R} \approx 10 f_{-4}$.  So $f_{-4} = 1$
represents a larger mass loss rate than one would get
assuming a typical value for normal red giants of $\eta_{\rm R} \approx 0.5$.
The \citet{Bloecker1995a} mass loss prescription 
for asymptotic giant branch stars
depends more steeply on $L$ and $M$,
such that with $L \approx \unit[3\times10^4]{\Lsun}$, $M \approx \unit[1.5]{\Msun}$,
and a typical scaling factor of $\eta_{\rm B} \approx 0.05$, it
yields a factor of $\approx 250$ enhancement over the Reimers prescription.
So a value $f_{-4} = 1$ represents a lower mass loss rate than
if we were to simply assume a Bl\"{o}cker prescription.

\begin{figure*}
  \centering
  \includegraphics[width=0.49\textwidth]{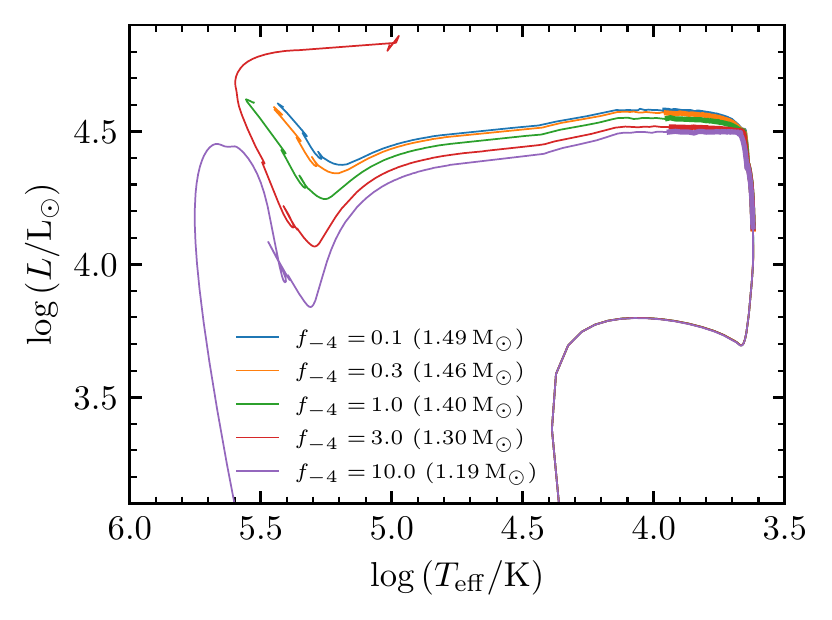}
  \includegraphics[width=0.49\textwidth]{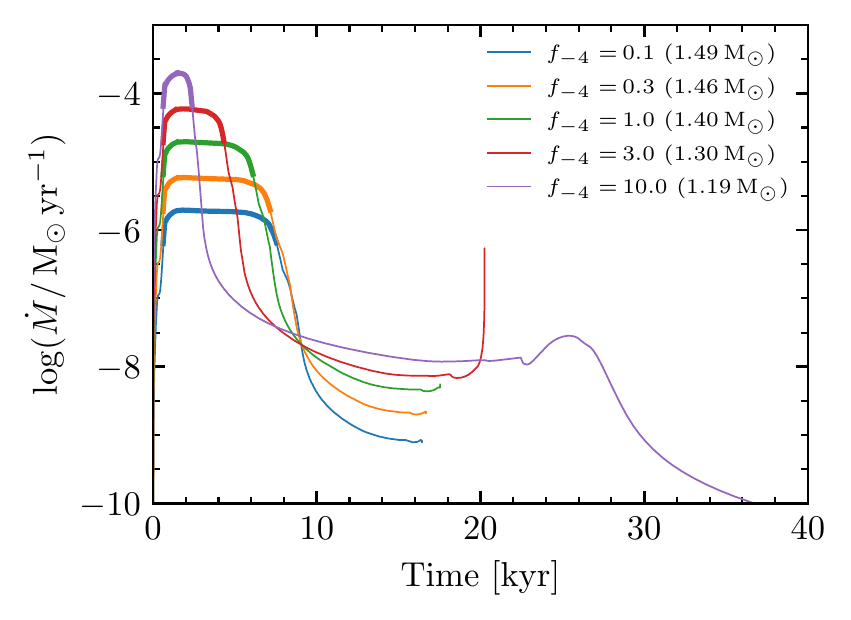}
  \caption{Comparison of the $q=2/3$, $M_{\rm tot} = 1.5\,\Msun$ model
    with varying mass loss rates using the prescription given by
    Equation~\eqref{eq:mdot-kh}.  The left panel shows the evolution
    in the HR diagram.  The right panel shows the mass loss rate as a
    function of time.  Most of the mass is lost in the vicinity of the
    luminous/cool corner of the evolutionary tracks; the thick portion
    of the matching lines in each panel indicate corresponding time
    intervals.  The legend parenthetically indicates the mass at the
    end of the calculation. \label{fig:compare-mdots}}
\end{figure*}

Figure~\ref{fig:compare-mdots} shows the HR diagram evolution of the
$q=2/3$, $M_{\rm tot} = 1.5\,\Msun$ model with different scaling
factors. The initial thermal adjustment phase is rapid, and so the
differences only appear as the remnant reaches its longer-lived
luminous giant phase and then begins to evolve to the blue.
The model with the highest mass loss rate ($f_{-4} = 10$) goes down
the cooling track as a $\approx \unit[1.2]{\Msun}$ ONe WD.  The model
with a somewhat lower rate ($f_{-4} = 3$) shrinks to
$\approx \unit[1.3]{\Msun}$ and undergoes a similar excursion as the
initially $\unit[1.3]{\Msun}$ model shown in
Figure~\ref{fig:HR-variation} before halting with the same numerical
problems. If the evolution continued and the NeO-burning successfully
reached the center, this might leave a WD with a Si-group core
composition.  The other three models with lower mass loss rates all
remain $> \unit[1.35]{\Msun}$, experience off-center Ne ignition, and
would likely go on to form low mass Fe cores and subsequently collapse
to NSs.  This illustrates how the amount of mass shed as a giant
qualitatively alters the evolution and that the mass as the object
leaves the giant phase controls the extent to which it undergoes
advanced burning stages and the type of object it leaves behind.

We also ran a set of models with constant mass loss rates during the
giant phase.  Tracks with constant mass loss rates comparable to the
values realized in the right panel of Figure~\ref{fig:compare-mdots}
gave qualitative agreement in terms of the HR tracks and final
outcomes.  This shows that the evolution is not sensitive to the
details of the mass loss prescription.  We expect that prescriptions
that give similar total mass loss over the giant phase will result in
similar evolution.

All of the material lost during these phases has the initial
composition of the WD.  The carbon burning is ignited relatively deep
in the remnant (around the temperature peak) and its neutrino-cooled
convection zone does not extend to near the surface.
Later, carbon flashes process material further out, and eventually only a
CO surface layer of mass $\sim 10^{-2}\,\Msun$ remains.
For representative Kippenhahn diagrams see Figures 2 and 5 in
\citet{Schwab2016b}.

\section{Effects of Nuclear Burning}
\label{sec:nuclear-burning}

\subsection{Helium burning}

The presence of He surface layers on the CO WDs allows for the
possibility of surface detonations and hence also double detonation
scenarios for thermonuclear supernovae
\citep[e.g.,][]{Dan2015}.  Our approach is not suitable for cases where
significant nuclear energy is released on the dynamical or viscous
timescales.%
\footnote{See Figure 12 in \citet{Dan2014}
  for the minimum burning timescales in their models. In the
  equipotential-averaged version of their models that guide our
  initial conditions, the minimum burning timescales are
  $\gtrsim 10^5$ s.}
Rather, we consider the effects of nuclear energy release
on the $\sim$ kyr evolutionary timescale of the remnants.

In \citet{Dan2014}, their HeCO WDs have masses 0.50, 0.55, and 0.60
\Msun\ and have pure CO cores overlaid by a $\unit[0.1]{\Msun}$ He
mantle.  This is a typical amount of He found on lower mass CO WDs in
stellar evolution calculations \citep[e.g.,][]{Zenati2019a}.  Thus for
mergers involving at least one lower mass CO WD
($\approx 0.5-0.7\,\Msun$), initially $\approx 10\%$ of the remnant
mass may be He.

When the lower mass WD is tidally disrupted, its composition is mixed
as it forms the disk/envelope.  The outer layers of the primary WD are
also mixed via the dredge-up action of the merger \citep{Staff2018}.
The chemical profiles from the \citet{Dan2014} models are generally
well-mixed, such that the He distribution can be reasonably
approximated as a constant value of $X_{\rm He}$ in the disk/envelope
component.  (The He mass fraction is higher in the outermost layers
reflecting material stripped early in the merger and now at larger
radii, including the tidal tail.)

The presence of a significant amount of He suggests that the object
will set up a steady He burning shell.  This is similar to what
happens in the formation of \mbox{R CrB} stars formed via He+CO WD
mergers, with the main difference being that the burning shell is
processing an envelope that is majority CO with a $\approx 10\%$ He
mass fraction, as opposed to the R CrB envelopes which are mostly He
with percent-level C abundances \citep[e.g.,][]{Asplund2000}. Given
that CO core WDs are O-dominated, one other conspicuous difference may
be a C/O number ratio $< 1$, suggesting an O-rich surface chemistry in
the cool outer layers.

This steady shell burning will extend the time the object spends in
the giant phase.  The specific energy associated with He burning is
$Q_{\rm He} \approx \unit[7\times 10^{17}]{erg\,g^{-1}}$.  Given core
masses $\gtrsim \unit[0.6]{\Msun}$, the shell burning luminosity is
expected to be a significant fraction of the Eddington luminosity
\citep[e.g.,][]{Jeffery1988, Saio1988b}.  Thus, assuming all the He
burns, the associated timescale is
\begin{equation}
  t_{\rm giant} \sim \unit[40]{kyr}
\left(\frac{L}{\unit[3\times10^4]{\Lsun}}\right) \left(\frac{M_{\rm He}}{\unit[0.1]{\Msun}}\right)~.
\end{equation}
That is similar to the lifetime of an R CrB star, reflecting the
similar luminosities and amount of He.  As discussed by
\citet{Schwab2019} in the R CrB context, if the mass loss rate becomes
comparable to the rate at which mass is processed through the He
burning shell, then the lifetime is limited by the removal of the
reservoir of He in the envelope.

\begin{figure*}
  \centering
  \includegraphics[width=0.49\textwidth]{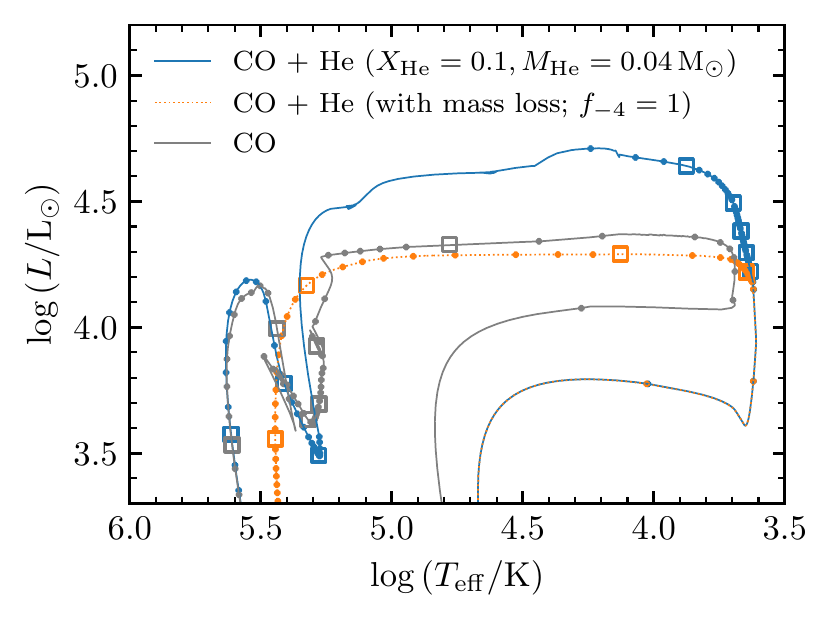}
  \includegraphics[width=0.49\textwidth]{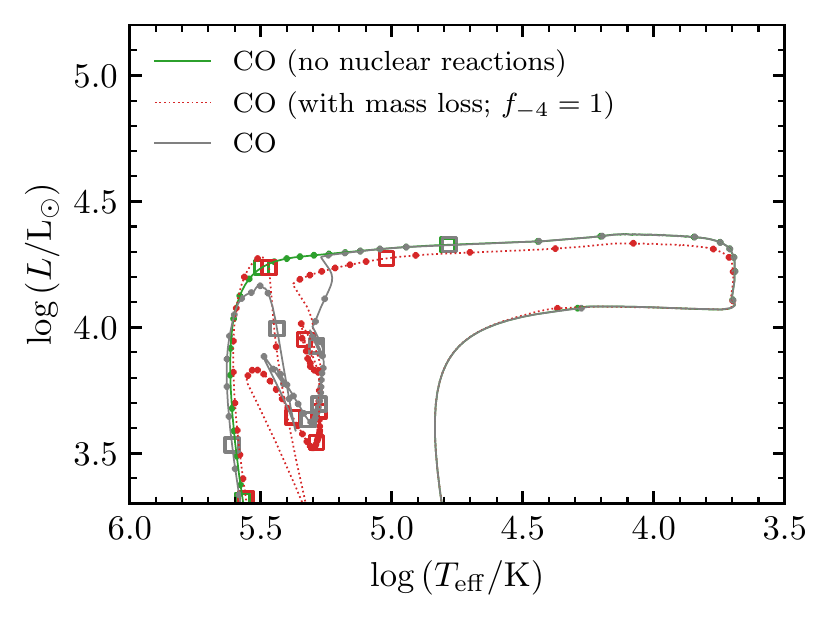}
  \caption{Evolution in the HR diagram of a $q=0.90$,
    $M_{\rm tot} = 1.1\,\Msun$ remnant.  Initial model ``CO'' has the
    default pure carbon-oxygen composition.  Initial model ``CO + He''
    has the indicated amount of He uniformly distributed in the
    disk/envelope material.  Solid lines show models without mass
    loss; dotted lines show models that include mass loss.  A small
    dot appears along the track each kyr and a large square each 10
    kyr. The left panel illustrates that the presence of He and the
    energy release from He burning extends the time spent as a giant
    by $\sim 10$ kyr.  The right panel illustrates the limited
    influence of carbon burning during the giant phase.  (See text for
    more discussion.) \label{fig:He-comparison}}
\end{figure*}

To illustrate the difference in evolution, we set $X_{\rm He} = 0.1$
in the outer layers of the $q=0.90$, $M_{\rm tot} = 1.1\,\Msun$ model.
This adds a total mass of 0.04\,\Msun\ of He.  This initial model is
indicated as ``CO + He'', while the model without He is indicated as
``CO''.  Figure~\ref{fig:He-comparison} compares the evolutionary
tracks in the HR diagram.  The reference (both panels, solid, grey
line) is the pure CO model without mass loss.
In the run without mass loss (left panel,
solid, blue line), the model with He spends approximately 10 times
longer than the ``CO'' model (40 kyr vs. 4 kyr) in the giant phase.
The increased luminosity and longer giant phase suggest that the
presence of He will make mass loss an even more important effect.  In
a run of the ``CO + He'' model with mass loss (left panel, dotted,
orange line), the object spends only $\approx 15$ kyr as a giant and
does not reach such extreme luminosities.

\subsection{Carbon burning}

Figure~\ref{fig:He-comparison} we show the case (right panel, solid,
green line) of the ``CO'' model in which nuclear reactions are not
included.  The evolution on the right of the HR diagram is the same,
reflecting that this phase radiates the thermal energy deposited in
the merger, not nuclear energy from carbon burning.
On the left of the HR diagram, the case without nuclear reactions
simply goes down the cooling track as a massive CO WD.  This track
notably lacks the drop in $L$ around $\logTeff = 5.3$ that is present
in other models.  This feature is associated with carbon ignition (see
marked point in Figure~\ref{fig:HR-senv}), the propagation of the
carbon burning to the center, and subsequent of carbon flashes in the
outer layers.  Therefore, tracks with this feature that also go down
the cooling track leave behind an ONe WD.

Note that in the case with He burning and mass loss (left panel,
dotted, orange line), this feature is absent.  The mass loss led to
less compression around the location of the temperature peak and a
long-lived carbon-burning front did not form.  The remnant in this
case is a $\unit[0.91]{\Msun}$ CO WD (with
$\unit[3\times10^{-4}]{\Msun}$ of He). By contrast, the ``CO'' model
evolved with the same mass loss prescription (right panel, dotted, red
line), but lacking the He-burning-associated mass loss, reaches the
cooling track as an $\unit[1.06]{\Msun}$ ONe WD.

\begin{figure}
  \centering
  \includegraphics[width=\columnwidth]{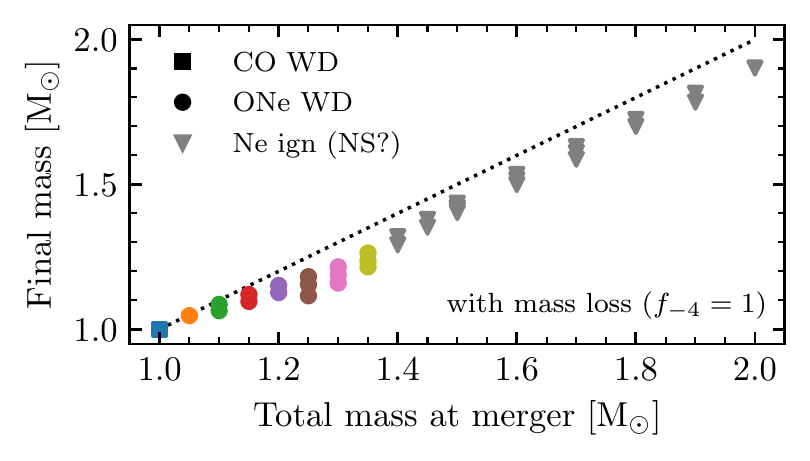}
  \caption{Final masses (and remnant type) for a set of models with
    varying total mass and mass ratio.  The dotted line marks
    conservative evolution (final mass equals initial mass). Since our
    models stop at Ne ignition, we cannot definitely say these form
    NSs, but that outcome seems inevitable for anything that remains
    super-Chandrasekhar (i.e., all but the lowest mass set of grey
    triangles, which might instead leave an Si- or Fe-group core
    WD). \label{fig:final-masses}}
\end{figure}

This illustrates that variations in the mass loss rate affect whether
particular advanced nuclear burning stages occur.  If we vary the mass
loss rate scaling factor for the $q=0.9$, $M_{\rm tot} = 1.1\,\Msun$
model, the $f_{-4} = 10$ case leaves behind a $\unit[1.02]{\Msun}$ CO
WD, whereas the $f_{-4} = 3$ case leaves a $\unit[1.04]{\Msun}$ ONe
WD.  Our models change whether or not they experience C-ignition
within the (post-giant-phase) mass range
$\approx \unit[1.0-1.1]{\Msun}$.  Figure~\ref{fig:compare-mdots} shows
the similar change for Ne-ignition occurring within a mass range
$\approx \unit[1.3-1.4]{\Msun}$.  Figure~\ref{fig:final-masses}
illustrates how the outcome varies with the initial total mass at
merger and the final remnant mass.  These transition masses are not
dissimilar from the typical characteristic masses in single star
evolution.

% higher mass ratio has higher total mass

\subsubsection{Implications for WD core compositions}

In single stars, the transition from CO WDs to ONe WDs is marked by
the occurrence of off-center carbon ignition in the degenerate,
neutrino-cooled CO core.
The characteristic core mass for this process to occur was calculated
by \citet{Murai1968}, who performed calculations of contracting CO
cores and found a minimum mass for carbon ignition of
\unit[1.06]{\Msun}.
This early estimate matches the maximum mass of
$\approx \unit[1.06]{\Msun}$ found in modern calculations of single
star evolution \citep[e.g.,][]{Doherty2015}.

The thermal structure of the merger remnants is not identical to the
CO cores in single stars, but is similar, with an off-center
temperature peak above a degenerate core.  Carbon burning is not
directly ignited by the merger in the lower mass merger remnants
relevant for understanding the CO / ONe transition.  Rather, the
compressional heating at the base of the cooling envelope leads to
carbon ignition.  This compression-induced ignition has long been
understood from work treating the WD-WD merger as the
Eddington-limited accretion of the secondary onto the primary
\citep{Nomoto1985, Saio1985b}. Within this framework,
\citet{Kawai1987} found ignition in CO WDs to occur at a mass
\unit[1.07]{\Msun}.  While we argue against the details of the
accretion picture, the compression at the base of a cooling envelope
radiating at the Eddington luminosity is similar to the compression at
the base of the accreted layer in an object accreting at the Eddington
rate \citep{Shen2012}.  Consistent with this understanding, our models
that do not experience carbon ignition are those that either begin at
masses $\lesssim \unit[1.05]{\Msun}$ or reach these lower masses
through significant mass loss.

\citet{Cheng2019} present evidence from \textit{Gaia} kinematics that
a population of massive ($\gtrsim \unit[1]{\Msun}$) WDs experiences
multi-Gyr cooling delays.  The location of these objects on the
Q-branch in the color-magnitude diagram (CMD) is coincident with the
expected location of crystallization for CO-core WDs
\citep[see also][]{Tremblay2019a}.  \citet{Bauer2020} argue
that this delay is explainable by an enhanced rate of \neon[22]
sedimentation in strongly liquid material near the liquid-solid phase
transition, providing further evidence that this sequence coincides
with the location of core crystallization.
Because of the higher mean charge in the plasma, WDs with ONe cores
crystallize at higher temperatures and thus at locations in the CMD
incompatible with the population identified by \citet{Cheng2019}.
Therefore, the observed Q-branch sequence appears to indicate the
presence of CO-core WDs with $\gtrsim \unit[1.2]{\Msun}$.
Based on our models, such objects continue to be surprising, as the
production of ultra-massive CO WDs is not a natural prediction of the
WD-WD merger scenario.

\section{Effects of Rotation}
\label{sec:rotation}

The orbital angular momentum of the WD binary at the point of tidal
disruption becomes part of the rotational angular momentum of the
merged object.  If the subsequent evolution of the remnant conserved
total angular momentum, the object would reach break-up as it
contracted towards a compact configuration \citep{Gourgouliatos2006}.

Compact configurations do become accessible so long as the evolution
is non-conservative \citep[see e.g., Figure 14 in][]{Yoon2007}.
During the viscous phase, angular momentum is
transported outwards.  As the object expands to its giant phase, its
angular momentum no longer need imply particularly rapid rotation.  As
the rotating outer layers of the envelope are shed, they can
dramatically reduce the total angular momentum.  For example, in the
\citet{Schwab2018} models of He WD + He WD merger remnants evolving to
form hot subdwarfs, 99\% of the angular momentum is lost by the
removal of 1\% of the mass.

The hot, differentially-rotating merger remnant
offers the opportunity to produce a strong, long-lived magnetic
field.  Simple equipartition arguments suggest the possibility
of fields $\sim \unit[10^{10}]{G}$ and 
\citet{GarciaBerro2012} show that dynamo action can easily produce fields of $\sim \unit[10^{7}]{G}$ under these conditions.

The magnetization and remaining angular momentum may play an important role in the
signatures and final fate of the remnant.  If the core remains rapidly
rotating, its spin-down might power a wind \citep[][]{Gvaramadze2019b, Kashiyama2019}.
For those remnants that evolve towards a core collapse event, the
rotation could play a role in the supernova and its explosion and/or
lead to the formation of a magnetar.
For those objects that leave behind massive WDs, the rotation may
provide a clue to their origins.  Since WDs are typically slow
rotators, rapid WD rotation (and strong magnetization) has often been interpreted as evidence for
a merger event \citep[e.g.,][]{Ferrario1997, Reding2020}.

\citet{Yoon2007}, who evolve stellar models of remnants with initial
conditions based on WD-WD merger simulations, include the effects of
rotation in their models.  They incorporate internal angular
momentum transport based on hydrodynamic (but not MHD) processes
and study the effect of varying a parameterized
angular momentum loss timescale.  They find that when this timescale is
longer than the neutrino cooling at the merger interface, rotation
prevents the compression of these initially hot layers.  This avoids
off-center carbon ignition and (in the super-Chandrasekhar case) leads
instead to central carbon ignition and a Type Ia SN.
However, the
inclusion of transport via a turbulent viscosity suggests compression
on a viscous timescale of hours to days as recognized by
\citet{LorenAguilar2009} and \citet{vanKerkwijk2010} and placed more
securely in an MHD context by \citet{Shen2012} and \citet{Ji2013}.  
As the \citet{Yoon2007} scenario for avoiding off-center
carbon ignition requires transport/loss timescales many orders of
magnitude slower than the MHD-motivated processes considered here, we do
not expect rotation to significantly modify the carbon burning
scenarios outlined in Section~\ref{sec:nuclear-burning}.

We construct a series of \MESA\ models including the effects of
rotation.
Similar our approach to the thermal structure of our models, we make
approximate choices for the initial angular momentum profiles.  We
then adopt various prescriptions for the internal angular momentum
transport in the remnant and the rate of mass loss from the surface.
We cap the rotational corrections to the structure \citep[Section
4,][]{Paxton2019} at those corresponding to 0.6 of critical rotation.
Some material in the envelope may exceed this value and in some cases
even become super-critical.  The evolution of this material cannot be
followed with any fidelity in our 1D calculations.
Nonetheless, our models provide a schematic picture of the rotational
evolution.

\subsection{First thermal time after merger}

In the simulations of \citet{Schwab2012}, the core spun down
substantially during the $\sim \unit[10^4]{s}$ viscous phase. However,
the simulations in \citet{Schwab2012} assume a constant $\alpha$ value
for the viscosity of $\sim 10^{-2}$.  This is reasonable for regions
where the transport is induced by the magneto-rotational instability
(MRI), but is likely an overestimate in the MRI-stable regions where
the operative viscosity is a less efficient process such as the
Tayler-Spruit dynamo \citep{Tayler1973, Spruit2002}.  While less
efficient than the MRI, we still expect that the core slows on
timescales much shorter than the kyr evolutionary timescale.
\citet{Shen2012} estimate a viscous timescale $\sim \unit[10^8]{s}$
associated with Tayler-Spruit.

To illustrate that we expect the core to no longer be rapidly rotating
by the time the object expands, we run a rotating \MESA\ model of the
$q=2/3$, $M_{\rm tot} = 1.5\,\Msun$ merger remnant with two versions
of transport due to the Tayler instability.  One is the \MESA\ version
of the Tayler-Spruit dynamo described in \citet{Paxton2013}, which we
refer to as ST.  This is based on the implementation of
\citet{Petrovic2005a} and \citet{Heger2005b}.  The other is the
treatment of the Tayler instability from \citet{Fuller2019a}, using
the authors' own implementation, which we refer to as FPJ.%
\footnote{We thank Adam Jermyn for making these routines available.}

We choose an initial core rotation profile to resemble the angular
velocities from the \citet{Dan2014} calculations (where the WDs were
assumed to be tidally locked) at the end of the dynamical phase.  This
makes the limiting assumption that no spindown occurred during the
elided viscous phase.  For the model we show,
$\Omega_0 \approx 0.35\,\rm s^{-1}$.  We place the rest of the remnant
at a fixed fraction of critical rotation, in this case chosen to be
$\approx 0.25$, as this gave a total angular momentum of the remnant
approximately equal to that from the merger calculation.

\begin{figure}
  \centering
\includegraphics[width=\columnwidth]{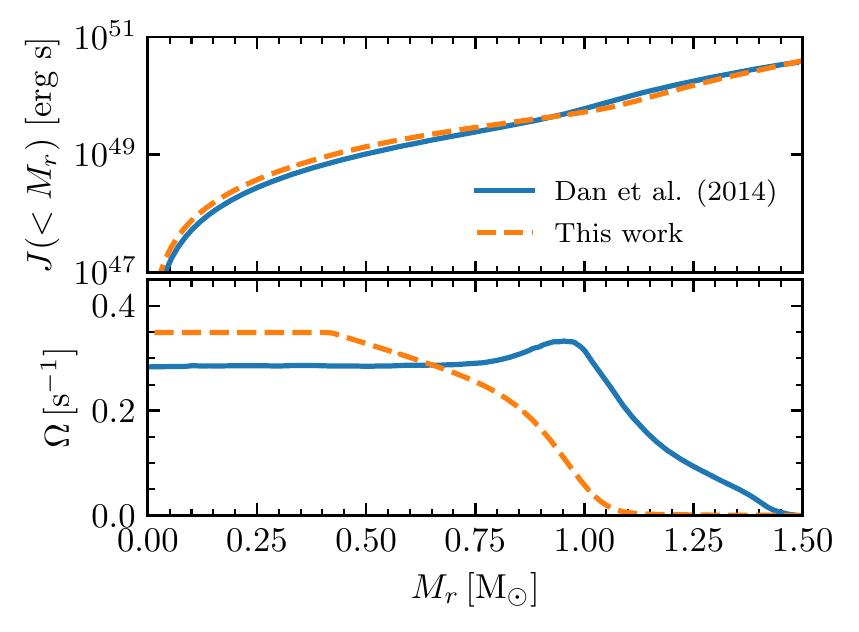}
\caption{Initial rotation properties for the $q=2/3$,
  $M_{\rm tot} = 1.5\,\Msun$ \MESA\ model compared to the immediate
  post-merger rotational properties of the analogous
  (equipotential-averaged) calculation from \citet{Dan2014}. The upper
  panel shows the total enclosed angular momentum at each mass
  coordinate. The lower panel shows the angular velocity.
  \label{fig:J-init}}
\end{figure}

Figure~\ref{fig:J-init} shows the initial rotation profile adopted in
the $q=2/3$, $M_{\rm tot} = 1.5\,\Msun$ \MESA\ model.  For comparison,
we show the immediate post-merger state of the analogous calculation
from \citet{Dan2014}.  The enclosed angular momentum profiles are
similar.  The difference in the structure of the outer hot envelope in
the \MESA\ model and the immediate post-merger state of the
\citet{Dan2014} calculation (see Figure~\ref{fig:ics-profiles} and
surrounding discussion) cause the more significant difference in
$\Omega$.  For our purposes, the important points of comparison are
that the total angular momentum is approximately the same and that the
cores have similar sizes and angular velocities.

\begin{figure}
  \centering
  \includegraphics[width=\columnwidth]{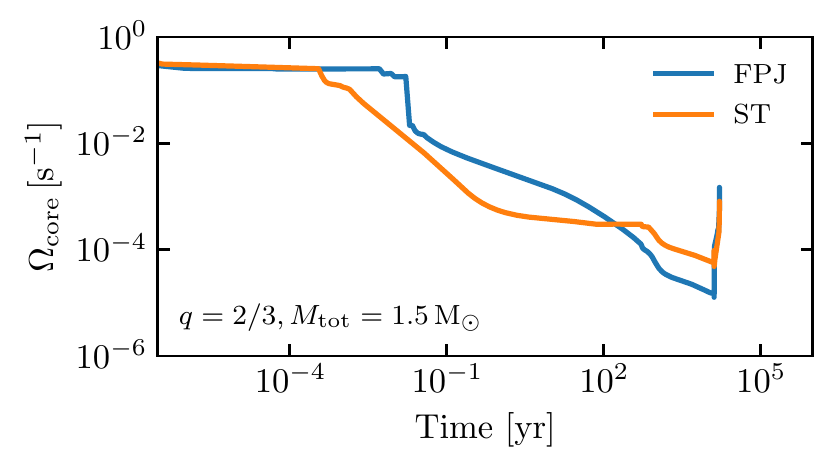}
\caption{Core rotation rate for the $q=2/3$,
  $M_{\rm tot} = 1.5\,\Msun$ model assuming two different schemes for
  angular momentum transport driven by the Tayler instability
  \citet{Tayler1973}. FPJ indicates the scenario and prescription of
  \citet{Fuller2019a}; ST is the \MESA\ implementation of the classic
  Tayler-Spruit dynamo \citep{Spruit2002}.  The core is initialized as
  shown in Figure~\ref{fig:J-init}.  In both cases, the core spins
  down on a timescale shorter than the thermal time in the remnant
  envelope, so this core spindown occurs before the object has reached
  the giant phase.  \label{fig:omega-core-early}}
\end{figure}

Figure~\ref{fig:omega-core-early} shows the time evolution of the core
rotation in a merger model using each of the ST and FPJ treatments.
These calculations confirm the essential point that we expect the
initially rapidly rotating core to spin down on $\sim$ yr timescales,
faster than the remnant can thermally adjust.%
\footnote{Because of this rapid spindown, our models do not predict a
  phase with a configuration matching the magnetic wind model used by
  \citet{Kashiyama2019} to interpret the possible WD-WD merger remnant
  reported by \citet{Gvaramadze2019b}.}
Therefore, we expect
that this spindown energy is deposited in very optically thick
material and so primarily goes into work to expand the remnant.
This spindown energy from the core will not be significant compared to
the energy thermalized during the dynamical and viscous phases of the
merger.  A rough way to understand this is that the merger thermalized
the kinetic energy of the orbit, the orbit and WD rotation have the
same period, but the orbital moment of inertia of the binary is much
greater than the rotational moment of inertia of the primary WD.

\subsection{Giant phase and beyond}

In the previous subsection, we argue that the core is unlikely to
remain rapidly rotating on kyr timescales.  On that first thermal
timescale, the remnant swells to become a giant and hence the envelope
also no longer need be rapidly rotating because of its large radius.

The remnant may have shed some angular momentum on the way to the
giant phase---though our models cannot reliably quantify the amount,
as it will depend on the details of the initial thermal and rotational
state of the outer layers and the prescriptions for removing mass.
During the giant phase, as discussed in Section~\ref{sec:mass-loss},
the remnant presumably loses some mass (and the accompanying angular
momentum).  However, if it retains even a few percent of the total
angular momentum at merger, it will still return to rapid rotation as
the envelope starts to cool and the object approaches a compact
configuration.

\begin{figure}
  \centering
\includegraphics[width=\columnwidth]{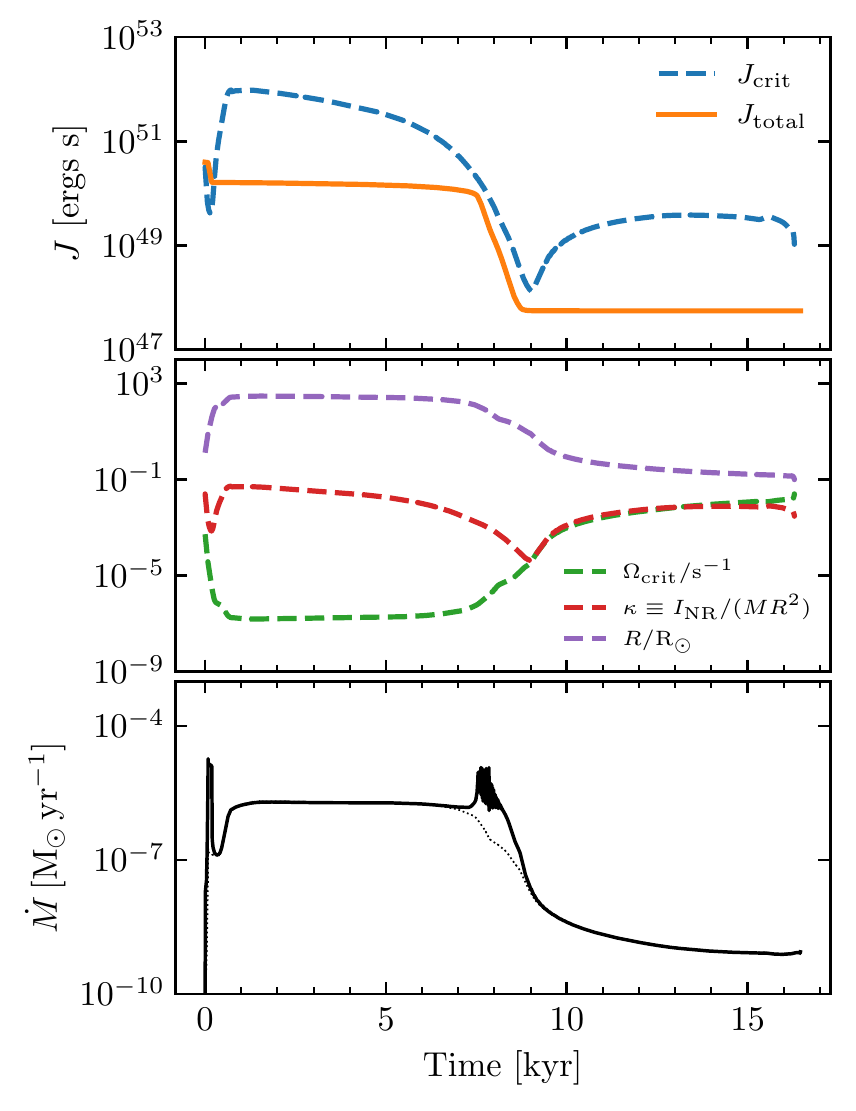}
\caption{Rotation-related quantities for the evolution of the $q=2/3$,
  $M_{\rm tot} = 1.5\,\Msun$ \MESA\ model using FPJ angular momentum
  transport.  Dashed lines show quantities from the non-rotating
  model.  In the upper panel, $J_{\rm crit}$ is the critical angular
  momentum, evaluated as
  $J_{\rm crit} \equiv I_{\rm NR} \Omega_{\rm crit}$. The middle panel
  shows these two values individually as well as the remnant radius.
  In the upper panel, $J_{\rm total}$ is the total angular momentum of
  the rotating model that sheds its critically-rotating outer layers.
  The lower panel shows the mass loss rate in this model.  The solid
  line shows the realized mass loss rate in the model while the thin
  dotted line indicates what the mass loss rate would be if
  critically-rotating material were not being
  removed.  \label{fig:J-evol}}
\end{figure}

From non-rotating models, we can make a simple estimate of the total
angular momentum that an object would be able to retain.  We do so by
calculating the angular momentum of the object if it were in solid
body rotation at the critical rotation rate of the surface:
$J_{\rm crit} \equiv I_{\rm NR} \Omega_{\rm crit}$, where $I_{\rm NR}$ is the
moment of inertia of the non-rotating model and
$\Omega_{\rm crit} = \sqrt{GM/R^3}$ is the critical rotation of the
surface.  (This simple estimate neglects the effects of rotational
deformation, the additional outward force from the near-Eddington
luminosity, and differential rotation.)  The dashed line in the top
panel Figure~\ref{fig:J-evol} shows this quantity as a function of
time in the $q=2/3$, $M_{\rm tot} = 1.5\,\Msun$ \MESA\ model.  This
quantity generally reaches a minimum as the object crosses to the blue
after the giant phase.  The middle panel shows the contraction of the
radius at this time and the evolution of the quantities that make up
$J_{\rm crit}$.

\begin{figure}
  \centering
\includegraphics[width=\columnwidth]{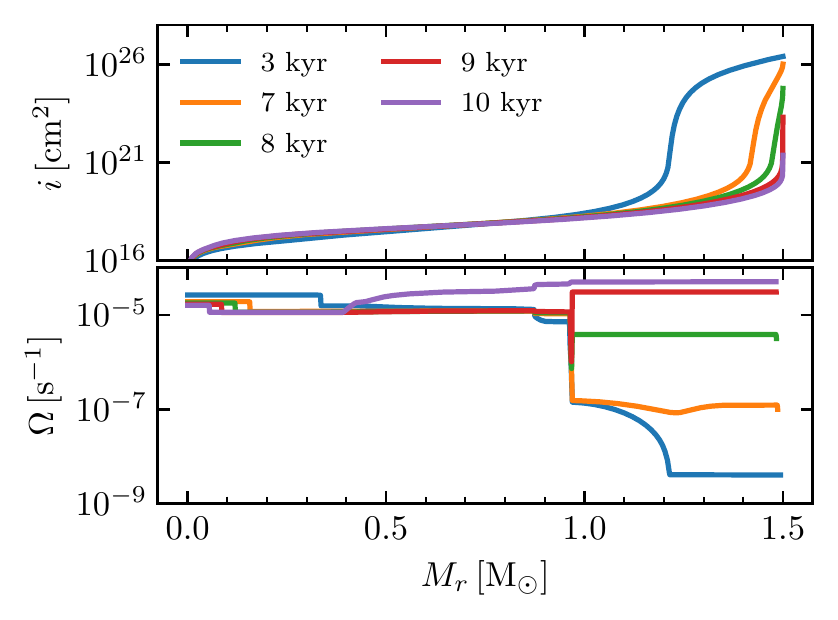}
\caption{Internal profiles of the models from Figure~\ref{fig:J-evol}
  at indicated times.  The top panel shows the specific moment of
  inertia.  The bottom panel shows the angular
  velocity.  \label{fig:I-Omega}}
\end{figure}

The solid line in the top panel of Figure~\ref{fig:J-evol} shows the
evolution of the total angular momentum in the rotating $q=2/3$,
$M_{\rm tot} = 1.5\,\Msun$ \MESA\ model with the FPJ angular momentum
transport prescription.  Angular momentum loss occurs such that
$J_{\rm total} < J_{\rm crit}$.  In addition to a small mass loss rate
($f_{-4} = 0.1$), we use built-in \MESA\ capabilities that can, at
each timestep, find the mass-loss rate that will keep the star below
critical rotation.  This approach avoids relying on any particular
form for rotationally-enhanced mass loss but still rapidly removes
super-critical material from the model.  During the evolution
$\approx \unit[0.02]{\Msun}$ of material is shed.  The lower panel of
Figure~\ref{fig:J-evol} shows the mass loss rate in the model as a
solid line.  The thin, dotted line shows what the mass loss rate would
be without this enhancement and there are two clear peaks 
corresponding to the periods when $J_{\rm total} \sim J_{\rm crit}$.
The amount of angular momentum lost during first kyr depends on our
initial conditions and so may not be reliable.  However, even if all
of the angular momentum were retained during this early phase, the
object will later shed angular momentum as it goes through the even
more restrictive (i.e., lower $J_{\rm crit}$) bottleneck (at 8-10 kyr
in Figure~\ref{fig:J-evol}).

The bottleneck occurs because the star is transitioning from having a
significant amount of mass in an extended envelope to a having a more
compact configuration.
The middle panel of Figure~\ref{fig:J-evol} shows the evolution of
the dimensionless moment of inertia $\kappa = I/(MR^2)$, along with $R$
and $\Omega_{\rm crit}$.  At fixed mass,
$J_{\rm crit} \propto \kappa \sqrt{R}$, and we see the value $R$
monotonically decreases, while the value $\kappa$ displays the
minimum.  When the star is a giant, there is a significant amount of
mass out in the envelope at a radius comparable to the surface radius.
Once the star has reached a compact configuration, again, a
significant amount of mass is at a radius comparable to the surface
radius.  However, in between, while the envelope is contracting, there
is a smaller fractional amount of mass near the surface.
Figure~\ref{fig:I-Omega} shows information about the internal
structure of the remnant at selected times.  The top panel plots the
specific moment of inertia $i$ (i.e., $I_{\rm NR} = \int i\,dm$).  The
contribution of the core remains constant (and sub-dominant), but the
contribution of material in the envelope
($M_r \gtrsim \unit[1.0]{\Msun}$) changes significantly, reflecting
the previously described transition.
The bottom panel of Figure~\ref{fig:I-Omega} shows the profiles of
$\Omega$.  With the assumed FPJ AM transport, the envelope generally
remains near solid body rotation.

Immediately post-merger objects have total angular momentum
$J \sim \unit[10^{50}]{erg\,s}$ well above the minimum $J_{\rm crit}$
they will reach in their subsequent evolution.  Motivated by
Figure~\ref{fig:J-evol}, we make the assumption that the objects that
evolve to the blue (either to collapse to an NS or to go down the WD
cooling track) have their total angular momentum reduced to a
characteristic value $J_{\rm final} \sim \min J_{\rm crit}$.  In our
models, $J_{\rm final} \sim \unit[10^{48}]{erg\,s}$.  Assuming that
subsequent evolution is conservative, this reflects the angular
momentum of the final remnant, and so we can estimate the associated
rotation periods.

\begin{figure}
  \centering
  \includegraphics[width=\columnwidth]{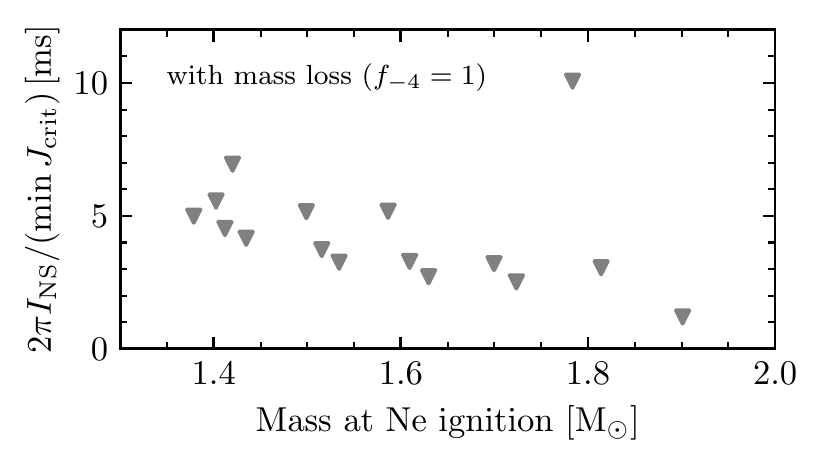}
  \caption{Estimated rotational periods of NSs.  Each symbol is one of the subset of models shown in Figure~\ref{fig:final-masses} that experience Ne ignition and seem likely to undergo an MIC to form a NS.  \label{fig:final-Prot-ns}.  Under the assumption of a conservative collapse, the x-axis would approximate the NS (baryonic) mass and the y-axis its rotational period.  This simple estimate assumes $I_{\rm NS} =  \unit[10^{45}]{g\,cm^2}$ with no mass dependence.}
\end{figure}

The moment of inertia of a NS is $I \sim \unit[10^{45}]{g\,cm^2}$,
implying $P_{\rm rot} \sim \unit[10]{ms}$.
Figure~\ref{fig:final-Prot-ns} shows
the rotational periods of the subset of models from
Figure~\ref{fig:final-masses} that likely collapse to NSs.
If the merger generates a $\sim \unit[10^9]{G}$ field that is then amplified by $\sim 10^4$ via flux conservation in the collapse, this could lead to the formation
of a millisecond magnetar.

\begin{figure}
  \centering
  \includegraphics[width=\columnwidth]{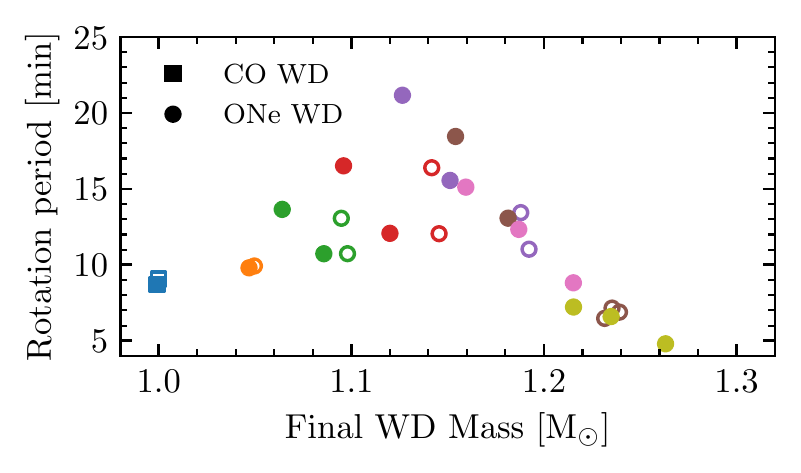}
  \caption{Estimated rotational periods of single WDs on the cooling
    track.  Each solid symbol is one of the subset of models shown in
    Figure~\ref{fig:final-masses} that leave single WDs.
    The color indicates the total mass at merger via Figure 9.
    Open symbols are a set of models with lower mass loss $f_{-4} = 0.1$.
    The rotational periods assume that the total angular momentum of the
    WD is the minimum value of the critical angular momentum
    encountered in the post-giant-phase evolution of the
    remnant. \label{fig:final-Prot}}
\end{figure}

The moment of inertia of a massive WD is
$I \sim \unit[10^{50}]{g\,cm^2}$, implying
$P_{\rm rot} \sim \unit[10]{min}$.  Figure~\ref{fig:final-Prot} shows
the rotational periods of the subset of models from
Figure~\ref{fig:final-masses} that leave WDs.\footnote{These models were
additionally cooled until $\log(L/\rm L_\odot) = -1$ so that the
moment of inertia would be closer to that of WDs observed on the
cooling track.}   The rotational periods are mostly in the range
$\approx \unit[10-20]{min}$,
with the most massive objects at $\gtrsim \unit[1.2]{\Msun}$
  having shorter periods of $\approx \unit[5-10]{min}$.
  The mass ratio is not explicitly indicated, but increasing mass ratio (at fixed total mass) typically gives higher final masses and shorter periods.
  We also plot a set of models with less assumed mass loss ($f_{-4} = 0.1$)
  as open symbols.  Since the AM bottleneck occurs after the giant phase where the mass loss is most important, the predicted rotational periods
  do not depend strongly on this choice.

Such periods are much shorter than typical WD rotational periods,
which are in the range of hours to days \cite[e.g.,][]{Hermes2017e}.
One class of atypical objects are the hot DQs \citep{Dufour2008}, many
of which have photometrically-detected periods (likely associated with
rotation) in the range $\approx \unit[5-20]{min}$
\citep{Williams2016}, though one object in this class does have a more
typical 2.1 d period \citep{Lawrie2013} and some do not have detected
variability.
A few individual objects with rapid rotation are also known.  RE
J0317-853 has a 12 min rotation period, a $\sim 100$ MG field, and a
mass $\approx \unit[1.3]{\Msun}$.  The case for this object as WD-WD
merger remnant is somewhat complicated by the fact that it is a DA WD
and in orbit with another WD with a roughly similar cooling age
\citep{Barstow1995, Ferrario1997, Vennes2003, Kuelebi2010}.
\citet{Reding2020} found the as yet fastest-rotating, apparently
isolated WD with a period of $\approx \unit[5.3]{min}$ and a mass
$\approx \unit[0.65]{\Msun}$.  Current and future ground- and
space-based photometric surveys are expected to enlarge our sample of
rapid WD rotators.  Our models suggest that a rotation period of
$\sim \unit[10]{min}$ in a single WD is a natural signature of its
origin in a WD-WD merger.

\section{Conclusions}
\label{sec:conclusions}

\begin{figure*}
  \centering
  \includegraphics[width=\textwidth]{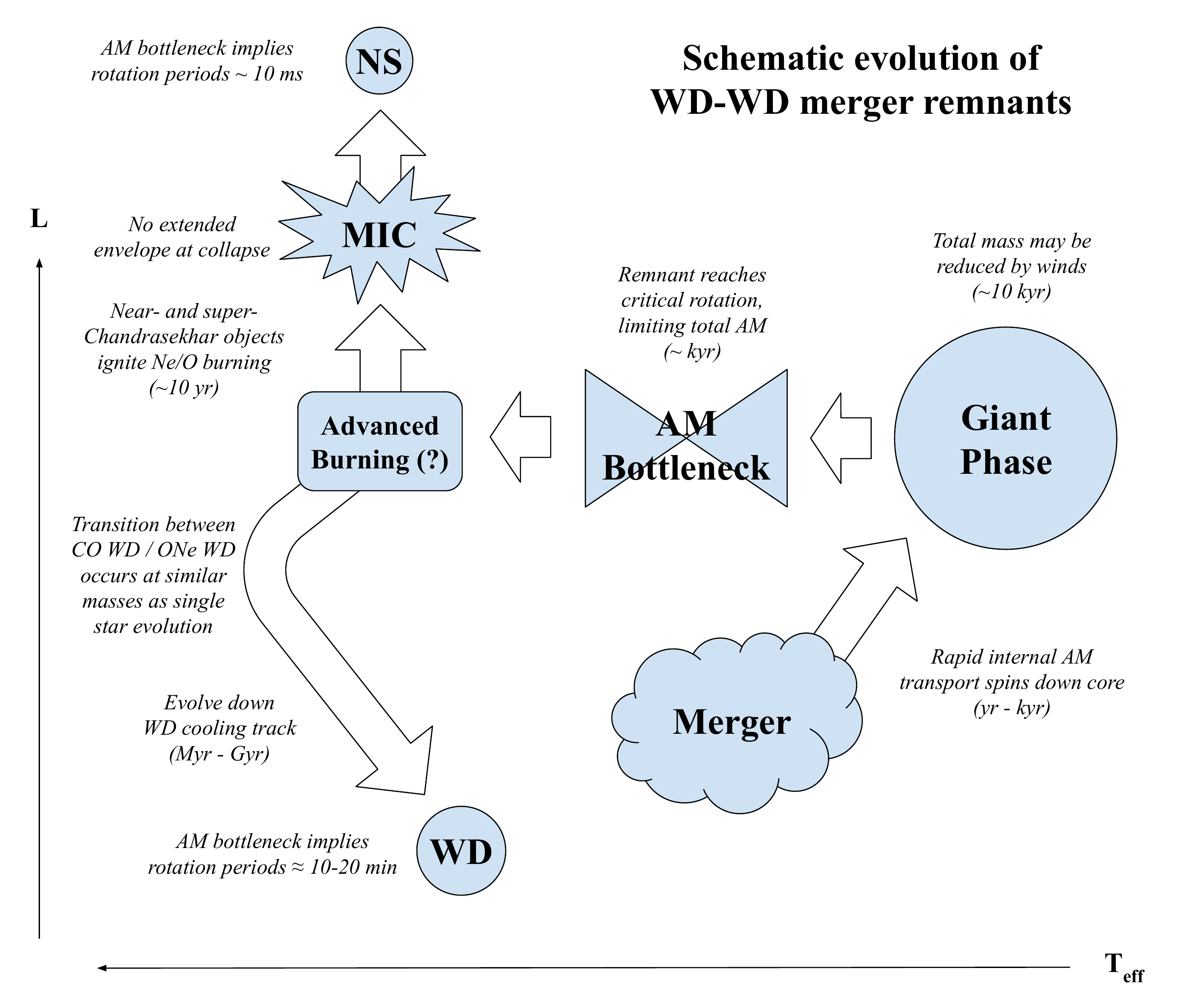}
  \caption{Evolutionary scenario outlined in this work illustrated on
    a schematic HR diagram.  Key phases and corresponding conclusions
    from this work are indicated.  \label{fig:schematic}}
\end{figure*}

We construct 1D stellar evolution models of the remnants of the merger
of two CO-core WDs using \MESA.  We use the results of hydrodynamic
calculations of the dynamical merger process from \citet{Dan2014},
along with the picture of the post-merger viscous phase developed in
\citet{Shen2012} and \citet{Schwab2012}, to construct approximate
initial conditions for merger remnants with a range of total mass and
mass ratios.  This allows us to survey the possible outcomes for these
mergers, extending beyond the single case considered in
\citet{Schwab2016b}, and paying increased attention to cases that
leave behind single, massive WDs.  This suite of models provides a
useful outline of the post-merger evolution.  We schematically
summarize this evolution and some of our conclusions in
Figure~\ref{fig:schematic}.

Immediately following the merger, the core of the remnant is rapidly
rotating ($\Omega \sim 0.3\,\rm s^{-1}$), reflecting the short orbital
period of the tidally-locked binary at merger.  When MHD angular
momentum transport process are included, we find (in agreement with
\citealt{Shen2012}), that the core spins down on $\sim$ yr
timescales, which is much less than the thermal time of the remnant.

On the $\sim$ kyr thermal time, the remnant evolves into a giant.  It
remains a giant for the $\sim 10$ kyr required to radiate away the
thermal energy deposited during the merger.  If significant He is
present on the WDs at merger (but not detonated), then a stable
He-burning shell is set up.  This energy release can extend the
lifetime of this phase by up to a factor of $\approx 10$, setting up
an object similar to an R CrB star, except with a He-deficient,
CO-dominated atmosphere.
Given the duration, if the Milky Way merger rate of appropriate WD-WD
systems is $\sim$ 1 per 300 yr, then we predict the existence of
$\sim 30$ galactic objects in this phase.

During the giant phase, the high luminosities, large radii, and metal-rich
surface suggest that significant mass loss can occur.  This mass loss
limits the duration of the giant phase and reduces the mass of the
remnant.  Simple prescriptions suggest that this is the most
non-conservative phase of the post-merger evolution, with the remnant
shedding $\sim \unit[0.1]{\Msun}$ of material.
By setting the remnant mass, this phase helps determine whether
advanced burning stages occur in the remnant interior and thus
influences whether the remnant becomes a NS or WD (and the core
composition of the WD).

Our models do not naturally yield ultra-massive CO WDs, that is WDs
with CO core compositions and masses significantly above the
$\approx \unit[1.06]{\Msun}$ minimum mass for ONe WDs formed in single
star evolution.  We find that off-center carbon ignition occurs and
converts the core to ONe unless the initial total mass is
$\lesssim \unit[1.05]{\Msun}$ or unless significant mass loss can
reduce the mass below this value.
We similarly find that Ne ignition occurs in remnants with total
masses $\gtrsim \Mch$ and that in the most cases, the remnant is in a
compact, blue configuration at the time collapse to an NS would occur
\citep[as also found in][]{Schwab2016b}.  Only in the most massive
remnant we consider (total mass $\approx \unit[1.9]{\Msun}$) does the
collapse to a NS occur in a still-inflated envelope of
$\approx \unit[30]{\Rsun}$.

As the remnant evolves blueward and the envelope contracts, we find
that an angular momentum ``bottleneck'' occurs.  The assumption of
solid body rotation and critical surface rotation define a
characteristic total angular momentum for the remnant and this
quantity reaches a minimum after the giant phase. Assuming that
minimum sets the amount of angular momentum retained by the remnant
beyond this point (and that further evolution is conservative), we
predict characteristic single WD rotation periods of
$\approx \unit[10-20]{min}$.  This strengthens the suggestion that
single WDs with these rotation periods are the products of WD-WD
mergers.  For the more massive cases that undergo a merger-induced
collapse,  NSs formed with this same amount of angular momentum would
have periods $\approx \unit[10]{ms}$.
These NSs may plausibly have magnetar-strength fields due to the
field generated in the aftermath of the merger and its subsequent enhancement during the collapse.  As such, WD-WD mergers provide an intriguing pathway for the formation of young magnetars in old stellar populations.

Future calculations that model the WD-WD merger and its immediate
aftermath will be of significant utility as they will provide
information about which systems experience thermonuclear explosions,
and, for the surviving systems, provide higher fidelity initial
conditions for the study of the remnant evolution

\acknowledgments

We are grateful to the anonymous referee and to Jim Fuller, JJ Hermes, Tony Piro, and Eliot Quataert for feedback that improved this manuscript.  We thank them and Evan Bauer, Lars Bildsten, Ilaria Caiazzo, Adam Jermyn, and Ken Shen for helpful conversations.
JS is supported by the A.F. Morrison Fellowship in Lick Observatory.
We acknowledge use of the lux supercomputer at UC Santa Cruz, funded by NSF MRI grant AST 1828315.

%\appendix

\clearpage

\bibliography{paper.bib}

\end{document}